\documentclass[prc,aps,preprint,showpacs]{revtex4}
\usepackage{graphicx}

\begin{document}
%\baselineskip 25pt minus .1pt
%\begin{center}
\title{One-particle properties of 
deformed N $\approx$ 28 odd-N nuclei with weakly-bound or
resonant neutrons} 

\author{ Ikuko Hamamoto$^{1,2}$ }

\affiliation{
$^{1}$ {\it Division of Mathematical Physics, Lund Institute of Technology 
at the University of Lund, Lund, Sweden}   \\
$^{2}$ {\it The Niels Bohr Institute, Blegdamsvej 17, 
Copenhagen \O,
DK-2100, Denmark} \\ 
}

%\end{center}

%\vspace{2cm}

%\date{\today}

\begin{abstract}
Possible deformation of odd-N nuclei with N $\approx$ 28 towards the neutron 
drip line is
investigated using the Nilsson diagram based on deformed Woods-Saxon potentials.
Both weakly-bound and resonant one-particle levels are properly 
obtained by directly 
solving the Schr\"{o}dinger equation in mesh of space coordinate 
with the correct boundary condition.
If we use the same diffuseness of the potential as that of 
$\beta$-stable nuclei, the energy difference between the neutron 
2p$_{3/2}$ and 1f$_{7/2}$ levels becomes very small or the N=28 energy gap
almost disappears, as the binding energies of those levels approach zero. 
This suggests that the ground states of those neutron drip line nuclei 
are likely to be deformed.
In particular, the spin-parity and the magnetic moment of the ground state 
of odd-N nuclei, $^{43}_{16}$S$_{27}$ and $^{45}_{16}$S$_{29}$, 
are examined. 
Moreover, it is suggested that in
$^{39}_{12}$Mg$_{27}$ lying outside the drip line the lowest resonant state may
have 5/2$^{-}$, if the N=28 energy gap almost vanishes.
\end{abstract}

\pacs{21.60.Ev, 21.10.Pc, 27.40.+z}

\maketitle

\section{INTRODUCTION}
The study of the properties of nuclei far from the line of $\beta$ stability is
presently one of the most active and challenging topics in nuclear structure. 
In particular, some neutron-rich nuclei 
with the traditional magic number N=20 are now 
known to be deformed.  It is theoretically expected that 
N=28 may also no longer be a magic number 
in some neutron-drip-line nuclei due to the behavior of
weakly-bound neutrons with $\ell$=1 relative to that with $\ell$=3, 
though very few experimental 
data are presently available for pinning down the change of this magic number.  
Keeping a model as simple as possible, in the present work 
we examine the change of the N=28 energy gap towards the
neutron drip line and its consequence, 
which is related to the characteristic feature of the 
weakly-bound and resonant one-particle levels with small $\ell$ values.  
This origin of the change of nuclear shell structure 
is quite different from the change related to the tensor force between
protons and neutrons using harmonic-oscillator wave functions.
The latter change of the neutron shell-structure depends on the number of
protons which occupy specific orbits and, furthermore, the effect of
weakly-bound neutrons is not taken into account.

The nucleus $^{40}_{12}$Mg$_{28}$ is known to lie inside the drip line, while 
$^{39}_{12}$Mg$_{27}$  lies outside \cite{TB07}.   It may still 
take a time to obtain 
spectroscopic informations on those neutron-rich Mg isotopes.
Whether or not a given neutron-drip-line nucleus 
is deformed depends also on the preference 
by the proton number.  In Ref. \cite{FS00} studying the 2n separation energy 
$S_{2n}$ as a function of the number of neutrons, 
it is pointed out that the S isotope exhibits a pronounced change of
slope around N=26, which is in contrast to the Ca isotope that is known to be 
spherical.
It is in general very useful to obtain the spin-parity of low-lying states of 
odd-N nuclei, which may directly provide the information on the neutron shell
structure.
In this sense, it is very interesting and important to pin down the spin-parity 
of neutron-rich nuclei, $^{43}_{16}$S$_{27}$ and $^{45}_{16}$S$_{29}$.
We refer to Ref. \cite{LAR08} for the present status of the study of the N=28
shell closure. 

In the present work we investigate the bound and resonant one-particle levels 
as a function of
deformation, using the parameters which may correspond to nuclei, 
$^{45}_{16}$S$_{29}$ and $^{39}_{12}$Mg$_{27}$.  
The change of neutron shell
structure at spherical as well as deformed shapes in the system with 
weakly-bound or resonant neutrons 
is studied. Calculated values of magnetic dipole moment for the neutron-rich 
odd-N S isotope 
are presented as an additional information, since the measured
magnetic moment gave a clear spin-parity assignment of 3/2$^{+}$ to the ground
state of $^{17}_{6}$C$_{11}$ with one weakly-bound neutron \cite{HO02} 
and 3/2$^{-}$
to that of $^{33}_{12}$Mg$_{21}$ \cite{DTY07}. 
The model used here is basically the same as that employed in Ref. \cite{IH07}.
It is the author's hope that the Nilsson diagrams of the present paper may be
useful as a basic help material in the interpretation of new experimental data 
on neutron drip line nuclei with N$\approx$28. 

In Sec. II the model is briefly described and discussed, 
while numerical results are presented in Sec.III.  
Conclusions and further discussions are given in Sec.IV.

\section{MODEL}
Though a considerable amount of publications are available for the study of 
possibly deformed N$\approx$28 nuclei towards the neutron drip line,  
to the author's knowledge no Hartree-Fock (HF) or Hartree-Fock-Bogoliubov (HFB) 
calculation for deformed nuclei 
is presently available which is carried out by integrating the Schr\"{o}dinger
equation in mesh of space coordinate with proper asymptotic behavior for
large $r$ values.  
This way of estimating one-particle levels is of essential importance, when we
are interested in the phenomena, in which 
resonant and/or weakly-bound
one-particle levels having small $\ell$-values as 
the major component of wave functions are 
playing an important role. 
Even if the technique to carry out that kind of deformed 
HF or HFB calculations is established, 
such a calculation gives one self-consistent
deformation which depends on the effective interaction used, and it is not easy 
to find out how general the obtained results are. 
Moreover, the formulation of traditional HF calculations is not directly
applicable for studying unbound nuclei such as $^{39}$Mg. 

In the analysis of experimental data on nuclei away from the stability line  
with N$\approx$20 and N$\approx$28 the prediction by the traditional shell model
with harmonic-oscillator wave-functions are so far often used as a convenient 
reference. 
The systematic shell-model calculation, in which
weakly-bound particles are properly treated, is indeed available \cite{VZ06}. 
However, the complicated calculations have so far been carried out only for 
very limited isotopes and certainly not for neutron-drip-line nuclei with
N$\approx$20 and N$\approx$28. 
If some nuclei are deformed or very soft in deformation, the
shell-model wave-functions become too complicated to extract the physics in a
simple intuitive terminology.  For example, the interesting
quantities such as one-particle energies and nuclear shape are not directly
obtained from shell model calculations.

Considering how useful the Nilsson diagram has been
in providing the basis for the classification of experimental data 
on stable deformed nuclei \cite{BM75}, 
in the present work we apply the model and idea presented in Ref.
\cite{IH07} to the study of neutron-drip-line nuclei with N$\approx$28.
In order to examine one-particle resonant states 
in the unbound nucleus $^{39}$Mg, in any case 
we have to assume a one-particle potential.
We use the parameters of Woods-Saxon potentials taken from the standard ones
\cite{BM69} for stable nuclei except for the depth, $V_{WS}$. 
Namely, the diffuseness, the strength of spin-orbit potentials and the radius
parameter are taken from those on p.239 of Ref. \cite{BM69}. 
A slightly larger diffuseness might be more appropriate for presently studied 
nuclei considering the contribution by weakly-bound neutron(s) to the
self-consistent potential.  However, 
we note that the major part of the nuclear potential is provided by well-bound
nucleons and, moreover, a larger diffuseness leads to the degeneracy of 
the 2p$_{3/2}$ and 1f$_{7/2}$ levels already 
at a larger binding energy than the one obtained in the present work.  
This is because for a given binding energy  
one-particle wave functions with small $\ell$ values can 
more easily extend to the outside of the potential, if the potential surface is
softer or more diffuse.  

The coupled equations derived from the Schr\"{o}dinger equation are 
solved in coordinate space with the correct asymptotic behavior of wave
functions for $r \rightarrow \infty$, both for bound \cite{IH04} and resonant
\cite{IH05,IH06} levels.  
The solution obtained in this way is totally independent of the upper limit of
radial integration, $R_{max}$, if both the potential and the coupling term 
are already negligible at $r = R_{max}$.

One-particle resonance is obtained if one of calculated 
eigenphases \cite{RGN66} increases through $\pi /2$ as energy increases. 
This is the definition of one-particle resonance in a deformed potential
\cite{IH05,IH06}, which is adopted in the present work.  The definition is a
natural extension of the definition of one-particle resonance for spherical
potentials in terms of phase shift \cite{RGN66}.  
In the numerical examples presented in this work where $A \approx 40$, 
for example, one-particle
resonant levels with $\Omega^{\pi}$ = 1/2$^{-}$ and 3/2$^{- }$ can hardly be
obtained for $\varepsilon_{\Omega} >$ 1.5 MeV 
if the main component of the wave functions inside the potential has $\ell$=1.
In contrast, well-defined 
resonant levels with $\Omega^{\pi}$ = 1/2$^{-}$ 
and 3/2$^{- }$ can be obtained up till several MeV 
when the major component inside the potential has $\ell \geq 3$.
See the dotted and dashed curves for $\varepsilon_{\Omega} > 0$ 
in Figs. 1 and 2.

\section{NUMERICAL RESULTS}
In Fig. 1 one-particle energies as a function of axially-symmetric quadrupole 
deformation are plotted, in
which $A$=45 and $V_{WS} = -41$ MeV are used.  The $V_{WS}$ value is
approximately equal to the value for $^{45}$S estimated from Eq. (2-182) 
of Ref. \cite{BM69}.  Considering that 
the measured neutron separation energy of $^{45}$S 
is 2.21 MeV, the level occupied by the 29th neutron has an approximately right 
binding-energy in both spherical and deformed cases.
In Fig. 1 one-particle resonant levels connected to the 1f$_{5/2}$ level 
at 1.90 MeV for $\beta$=0 
are plotted for all deformations as far as the resonance can be obtained 
in terms of eigenphase.  On the other hand, 
one-particle levels connected to the 1g$_{9/2}$ level are shown only around the
spherical shape, except
the $\Omega^{\pi}$ = 1/2$^{+}$ and 3/2$^{+}$ levels on the prolate side.  

It is seen that at $\beta$=0 the 2p$_{3/2}$ level lies 
2.27 MeV higher than the 1f$_{7/2}$ level.  The energy gap of 2.27 MeV is much
smaller than that for N=28 stable nuclei and may not be 
large enough to have N=28 as a magic number.
For comparison, the energy difference between 1f$_{7/2}$ and 1d$_{3/2}$, which
characterizes the magic number N=20 in Fig.1, is 4.92 MeV.
According to our experience in the analysis of stable nuclei in terms of Nilsson
diagrams, 
the ground state has a clear tendency to having the deformation 
where the local density of one-particle levels is minimum 
(or very small) for a given particle-number.  
It may further help to realize the deformation 
if some down-sloping one-particle levels around the Fermi
level are occupied compared with the configuration for the spherical shape.  
If we apply this rule, it is seen from Fig. 1 that the ground
state of N$\approx$28 nuclei with S$_{n}$ of a few MeV, namely
$^{43}_{16}$S$_{27}$ and 
$^{45}_{16}$S$_{29}$, 
may be prolate-deformed with $\beta$=0.3-0.5 rather than 
a spherical shape.
If so, the ground state of $^{43}$S with N=27 will have either 1/2$^{-}$ 
(dotted curve in Fig. 1) or 5/2$^{-}$ (dot-dashed curve in Fig. 1) while the
ground state of $^{45}$S with N=29 may 
have either 7/2${-}$ (dot-dot-dashed curve in
Fig. 1) or 1/2$^{-}$, since I$^{\pi}$ = $\Omega^{\pi}$ is expected for
respective band-head states in the region of $\beta$=0.3-0.5 
examining the one-particle 
wave functions in the Woods-Saxon potential.  
These values of spin-parity are quite different
from those for the spherical shape,
for which one expects 7/2$^{-}$ for the nucleus with N=27 and 3/2$^{-}$ 
for the nucleus with
N=29.   
Indeed, for example, the ground-state spin of all known N=27 isotones with even
Z$\geq$18 is 7/2$^-$, indicating that those nuclei are not far from being
spherical.

Next, we estimate possible values of magnetic dipole moment.
Either the p$_{3/2}$ or f$_{7/2}$ neutron in the seniority-one state produces 
the magnetic dipole moment of (1/2)$g_{s}^{eff}$, which is equal to 
$-$1.34 $\mu_{N}$ if we use $g_s^{eff}$ = (0.7)$g_s^{free}$. 
Indeed, the measured magnetic moment of the ground state (I$^{\pi}$=7/2$^-$) of
$^{47}_{20}$Ca$_{27}$ is $-$1.38(3) $\mu_{N}$ \cite{NJS05}, 
while the preliminary value of the ground state
(I$^{\pi}$=3/2$^-$) of $^{49}_{20}$Ca$_{29}$ is $-$1.38(6) $\mu_{N}$ 
\cite{NDS08}. 
The magnetic moment of deformed nuclei is calculated using 
Eqs. (4-86), (4-87) and (4-88) of Ref. \cite{BM75}, when the matrix elements of 
$\ell_{\nu}$ and $s_{\nu}$ are obtained from one-particle wave-functions in
the deformed potential (see Eqs. (5-86) and (5-87) of Ref. \cite{BM75}).
As an example of magnetic moments of deformed neutron-rich
nuclei, we may mention that the measured magnetic moment of the ground state
(I$^{\pi}$=3/2$^{-}$) of $^{33}_{12}$Mg$_{21}$ is $-$0.7456(5) $\mu_{N}$
\cite{DTY07}.  On the other hand, the value estimated in the present way is 
($-$0.88)-($-$0.75) $\mu_{N}$ for $\beta$=0.3 and
$g_s^{eff}$=(0.7-0.6)$g_s^{free}$, identifying the I$^{\pi}$=3/2$^{-}$ state as
the band-head state of the neutron [330 1/2] configuration (with the calculated 
decoupling parameter $a=-$3.2).

In Table I the calculated magnetic dipole moments using deformed 
one-particle wave
functions together with $g_{R}$=0.35 and $g_s^{eff}$ = (0.7)$g_s^{free}$ 
are shown.
Interestingly enough, calculated magnetic moments of $^{43}$S and $^{45}$S 
for the possible deformed shape are quite different from those 
for the spherical shape.  This suggests that the measurement of magnetic 
moments of the odd-N 
S-isotope may not only fix the spin-parity but also tell 
whether or not the nuclei are deformed.

Using the intermediate-energy Coulomb excitation of a radioactive $^{44}$S beam,
in Ref. \cite{TG97} the reduced transition probability B(E2;$0^{+}_{g.s.}
\rightarrow 2^{+}_{1}$) = 314(88) e$^{2}$ fm$^{4}$ 
with E$_{x}$(2$^{+}_{1}$) = 1297(18) keV
is obtained.  The relatively large B(E2) value is often quoted as an evidence
for a deformed shape of the nucleus $^{44}$S. 
Furthermore, in Ref. \cite{RWI99} a level at 940 keV with a B(E2) of
175(69) e$^2$ fm$^4$ in $^{43}$S is reported, also using the Coulomb excitation.
If we assume that the Z=16 proton core is the same for $^{44}$S 
and $^{43}$S, the B(E2) value is proportional to the Clebsch-Gordan coefficient
squared, [C(I$_i$ 2 I$_f$ ; K 0 K)]$^{2}$, 
for E2 transitions within a band with
a given K.  
Since the E2 excitation from the band-head state is here considered, 
the Coriolis perturbation may be neglected in the first approximation.  
Then, using the measured value of B(E2) in $^{44}$S, B(E2) values of
(I=K=5/2) $\rightarrow$ (I=7/2,K=5/2) and (I=K=1/2) $\rightarrow$ 
(I=3/2,K=1/2) in $^{43}$S are estimated to be 150(42) and 126(35) e$^2$ fm$^4$, 
respectively.  Both of these B(E2) values lie within the experimental error of 
the measured value of
175(69) e$^2$ fm$^4$.  Though the excitation energy of 940 keV seems to be 
a bit too large for the expected energy difference between the (I=K=1/2) and 
(I=3/2,K=1/2) states, the measured B(E2) value in Ref. \cite{RWI99} does not
really tell whether the possibly deformed ground state of $^{43}$S is 
I$^{\pi}$=K$^{\pi}$=5/2$^-$ or I$^{\pi}$=K$^{\pi}$=1/2$^-$.

In Fig. 2 the bound and resonant one-particle levels as a function of 
deformation are plotted for A=39 and V$_{WS}$=$-$37 MeV.  The parameters of the
Woods-Saxon potential are chosen approximately for the unbound nucleus 
$^{39}$Mg.  Though the $V_{WS}$ value estimated from Eq. (2-182) of 
Ref. \cite{BM69} is $-$38.3 MeV for $^{39}$Mg, a slightly shallower potential 
is used so that the one-particle level which the N=27th neutron will occupy 
is unbound for most of the likely deformation
including spherical shape.  The one-particle level scheme of Fig. 2 
may still give an N=28 system (such as $^{40}$Mg) 
inside the neutron drip line, since the pair correlation 
between the last-occupied pair of neutrons may produce 
an extra binding energy to the even-N system.

The 2p$_{1/2}$ level at $\beta$=0 and the $\Omega^{\pi}$=1/2$^{-}$
level connected to the 2p$_{1/2}$ level 
cannot be obtained as a resonant level and, thus, they do not appear  
in Fig. 2.  Both the $\Omega^{\pi}$=1/2$^{-}$ and 3/2$^{-}$ levels which are
connected to the 2p$_{3/2}$ level at $\beta$=0 cannot exist 
as resonant levels for 
$\varepsilon_{\Omega} > 1.25$ MeV, since the main component of the wave
functions inside the potential has $\ell$=1.  In contrast, the
$\Omega^{\pi}$=1/2$^{-}$ level connected to the 1f$_{5/2}$ level 
at $\beta$=0 can
continue as a resonant level for $\beta <$0.14 (or $\varepsilon_{\Omega} >$4.44
MeV) on the prolate side, since the major component of the wave function inside
the potential has then $\ell$=3.  For values of $\beta$ larger than 0.14 the
increasing portion of the $\ell$=1 component of the wave function 
inside the potential prevents the
eigenphase from reaching $\pi$/2. 

When the 2p$_{3/2}$ and 1f$_{7/2}$ levels approach zero binding, those two 
levels become almost degenerate.  See also Fig. 4 of Ref. \cite{IH07}.  
In Fig. 2 the distance between the 
2p$_{3/2}$ and 1f$_{7/2}$ resonant levels, namely the
energy gap at N=28, is only 180 keV.  
The partial filling of the almost degenerate neutron 2p$_{3/2}$ and 1f$_{7/2}$ 
shells may suggest a deviation from the spherical shape.  Moreover, the proton
number Z=12 (Mg-isotope) in the region of stable nuclei is known to prefer
prolate shape as seen from measured properties of $^{24}_{12}$Mg$_{12}$ and 
$^{25}_{12}$Mg$_{13}$.  Then, 
the spin-parity of the lowest-lying one-particle resonant
state of the unbound nucleus $^{39}$Mg is likely to be 5/2$^{-}$ 
for a possible appreciable amount of 
prolate deformation, examining the level structure around N=27 in Fig. 2.  
Indeed, the spin-parity of the level occupied by the 27th neutron 
on the prolate side depends
on the size of the N=28 energy gap at $\beta$=0, 
though on the oblate side it is always $\Omega^{\pi}$=1/2$^{-}$. 
For example, for an appreciable amount of prolate deformation 
the level occupied by the 27th neutron may have $\Omega^{\pi}$=1/2$^{-}$ 
in the case that the 2p$_{3/2}$
level lies appreciably higher than the 1f$_{7/2}$ level   
(see Fig. 3). 
If the lowest-lying one-particle resonant level in the possibly 
deformed (unbound) nucleus $^{39}$Mg is observed, the one-particle decay 
width will tell us whether
the resonance has I$^{\pi}$=5/2$^{-}$ or 1/2$^{-}$. 
At $\beta$=0 the 1f$_{7/2}$ resonance
at $\varepsilon$(1f$_{7/2}$) = 0.26 MeV has the calculated width of 0.8 keV,
while the 2p$_{3/2}$ resonance at $\varepsilon$(2p$_{3/2}$) = 0.44 MeV has 
the width of 890 keV.  Since the major component of the
$\Omega^{\pi}$=5/2$^{-}$ and 1/2$^{-}$ resonant levels inside the potential on
the prolate side has $\ell$=3 and $\ell$=1, respectively, the large difference 
of the widths of the I$^{\pi}$=5/2$^{-}$ and 1/2$^{-}$ levels remains unchanged
also for a moderate amount of prolate deformation.

Since some weakly-bound neutron orbits with $\Omega^{\pi}$=1/2$^{-}$ and 
3/2${^-}$ are occupied in the even-even core of $^{39}$Mg, one may wonder 
whether
the appropriate parameters of the Woods-Saxon potential given by the core in
this case may be different from those of stable nuclei.  In order to check the
validity of our numerical results of $^{39}$Mg we have tried the numerical
calculation by increasing the
diffuseness parameter $a$=0.67 fm of the potential 
used in Fig. 2 to $a$=0.75 fm.  
Then, we find the one-particle resonance energies 0.14 and 
0.37 MeV 
for the 2p$_{3/2}$ and 1f$_{7/2}$ neutron levels, respectively.
Namely, the 2p$_{3/2}$ resonant level lies now 
lower than the 1f$_{7/2}$ resonant
level by 230 keV.  Nevertheless, in the deformation region of 
0.25 $< \beta <$ 0.5 the negative-parity one-particle level scheme corresponding
to 20$<$N$<$28 remains nearly the same as in Fig. 2, because almost degenerate
2p$_{3/2}$ and 1f$_{7/2}$ levels strongly mix with each other as soon as 
deformation sets in.  Consequently, the statements on the unbound nucleus
$^{39}$Mg written in the previous paragraph are valid also for the larger
diffuseness $a$=0.75 fm.

For reference, in Fig.3 the Nilsson diagram for the case in which 
all levels of the 1f-2p shell at $\beta$=0 are well bound is shown.    
The parameters of the Woods-Saxon
potential are designed for the stable nuclei $^{52}_{24}$Cr$_{28}$.
The energy distance between the 2p$_{3/2}$ and 1f$_{7/2}$ levels at $\beta$=0, 
namely the energy gap at N=28,
is 4.03 MeV.

\section{CONCLUSIONS AND DISCUSSIONS}
First, being inspired by the observation that 
the behavior of $S_{2n}$ of the S isotope
for N$>$26 as a function of the number of neutrons exhibits a pronounced
difference from that of the Ca isotope, the possible spin-parity and the
magnetic moment of the nuclei $^{43}_{16}$S$_{27}$ and $^{45}_{16}$S$_{29}$
are estimated for an appreciable amount of prolate deformation.  
The study of the relevant Nilsson diagram indicates the possible prolate
deformation $\beta$=0.3-0.5 for those nuclei, applying the rule 
that a large one-particle level
density around the Fermi level at the spherical point may lead to a possible
deformation where the local one-particle level density is a minimum (or 
very small). 
For the possible prolate deformation; (a) The ground-state spin of $^{43}$S
($^{45}$S) is either 5/2$^-$ or 1/2$^-$ (7/2$^-$ or 1/2$^-$), in contrast to
7/2$^-$ (3/2$^-$) for the ground state of N=27 (N=29) spherical nuclei; 
(b) As shown in Table I, the calculated magnetic moments of the ground
states of $^{43}$S and $^{45}$S are clearly different from (1/2)$g_{s}^{eff}$ 
which is the value in the spherical limit of both N=27 and N=29.
Thus, it is very interesting to experimentally obtain any of the 
spectroscopic informations, in order to find an evidence of the change of the
magic number N=28.

For the ground state of the nucleus $^{44}$S a subtle competition between the
prolate and oblate deformations is previously obtained in the 
self-consistent mean field 
or more elaborate calculations by several theoretical groups.  
(See Ref. \cite{RRG02} and references quoted therein.) 
Examining the 
local density of one-particle levels around N=28 
in the Nilsson diagrams of Figs. 1, 2 and
3, it is seen that for a larger N=28 shell gap at $\beta$=0 such as in Fig. 3
the energy minima of prolate, spherical and oblate shapes may compete, while for
a smaller N=28 energy gap 
at $\beta$=0 the oblate shape is likely to be unfavored.  
In presently available HF or HFB or more elaborate calculations the
weakly-binding and resonant one-particle levels do not seem to be
properly treated.  Therefore, the size of the N=28 shell gap and the shell
structure of weakly-binding neutrons, which are obtained in those elaborate
calculations, need to be checked in a more careful way, though some of the
calculations certainly showed a reduction of the N=28 spherical
shell gap due to one or other reasons.  On the other hand,
protons are deeply bound in those neutron-rich nuclei and, thus, the 
deformation preferred by Z=16 protons may be 
guessed, in the first approximation, from the Nilsson diagram obtained by
using wave functions based on harmonic-oscillator wave functions.  For example, 
examining Figs. 2 and 3 of Ref. \cite{RRG02}, it is seen that 
the deformation preferred by Z=16 protons may be prolate rather than oblate.

Secondly, the spin-parity of the "ground" state (namely the lowest 
one-particle resonant 
state) of the unbound nucleus $^{39}_{12}$Mg$_{27}$ is investigated and
suggested to
be 5/2$^{-}$ if the N=28 energy gap almost collapses and 
the even-even core which provides the
potential is moderately prolate-deformed.  
In contrast, the lowest one-particle resonant state of $^{39}$Mg may have 
1/2$^-$ for a prolate-deformed core if the N=28 energy gap at the spherical
point is 
appreciable.
The expected spin-parity for the deformed shape makes a contrast to 
7/2$^{-}$ which is expected for a spherical nucleus with N=27.  
The resonant state with 5/2$^{-}$ or 7/2$^{-}$ where the minimum $\ell$ value of
the wave-function components $\ell_{min}$=3 can easily
be differentiated from the resonance with $\ell_{min}$=1 
if the one-particle decay 
width of the resonance is
observed.

In the present paper the possible many-body pair correlation is not included. 
However, we note that the spin-parity as well as the 
magnetic dipole moment of the ground state obtained in the present work 
will remain unchanged when one particle is replaced by one quasiparticle 
which comes from a mean-field approximation to the
many-body pair correlation.

The author is grateful to Professor P. Cottle for informative and stimulating
communications.

\newpage
\begin{table}[hbt]
\caption{Calculated magnetic dipole moments of the N=27th and 29th neutrons 
for the deformation $\beta$ = 0.35 and 0.45 in Fig. 1. Values of 
$g_{R}$=0.35 and
$g_{s}^{eff}$=(0.7)$g_{s}^{free}$ are used.  Corresponding one-particle levels 
in Fig. 1 are identified  
from the tabulated values of $\beta$ and $\varepsilon_{\Omega}$ and 
noting $I^{\pi}$ = $\Omega^{\pi}$.
The first and second rows correspond to the N=29th neutron in Fig. 1 depending 
on $\beta$, while the third and fourth rows to the N=27th neutron. 
}
\vspace*{2pt}

\begin{tabular}{|c|c|c|c|c|} \hline
%& & & &  \\
$I^{\pi}$ & $\mu$ ($\beta$ = 0.35) &
$\varepsilon_{\Omega}$ ($\beta$ = 0.35)  & 
$\mu$ ($\beta$ = 0.45) & $\varepsilon_{\Omega}$ ($\beta$ = 0.45) \\
 & ($\mu_{N}$) & (MeV) & ($\mu_{N}$) & (MeV)  \\ \hline

%& & & &  \\
7/2$^-$ & $-$0.765 & $-$1.09 & $-$0.765 & $-$0.38   \\
1/2$^-$ & +0.545 & $-$1.21 & +0.584 & $-$1.61  \\ 
5/2$^-$ & $-$0.580 & $-$4.19 & $-$0.600 & $-$4.47 \\
1/2$^-$ & +0.600 & $-$3.81 & +0.545 & $-$4.83  \\  \hline
%& & & &   \\ \hline

\end{tabular}

\end{table}

\newpage

\noindent
{\bf\large Figure captions}\\
\begin{description}
\item[{\rm Figure 1 :}]
Neutron one-particle levels as a function of axially-symmetric 
quadrupole deformation.
Parameters of the Woods-Saxon potential are designed approximately for 
the nucleus $^{45}_{16}$S$_{29}$.
The diffuseness, the radius and the depth of the Woods-Saxon potential are 
0.67 fm, 4.52 fm, and $-$41.0 MeV, respectively.
The $\Omega^{\pi}$ = 1/2$^{-}$ levels are denoted by dotted curves, the 
3/2$^{-}$ levels by dashed curves, the 5/2$^{-}$ levels by dot-dashed curves and
the 7/2$^{-}$ levels by dot-dot-dashed curves, while positive-parity levels are 
plotted by solid curves.
The neutron numbers 20 and 28, which are obtained by filling in all lower-lying
levels, are indicated with circles.
See the text for details.
\end{description}

\noindent
\begin{description}
\item[{\rm Figure 2 :}]
Neutron one-particle levels as a function of axially-symmetric 
quadrupole deformation.
Parameters of the Woods-Saxon potential are designed approximately for 
the unbound nucleus $^{39}_{12}$Mg$_{27}$.
The diffuseness, the radius and the depth of the Woods-Saxon potential are 
0.67 fm, 4.31 fm, and $-$37.0 MeV, respectively.
Neither the 2p$_{1/2}$ level at $\beta = 0$ nor one-particle levels at 
$\beta \neq$0 connected to the 2p$_{1/2}$ level are obtained 
as one-particle resonant levels and, thus, are not plotted.   
See the text for details and the caption to Fig. 1.
\end{description}

\noindent
\begin{description}
\item[{\rm Figure 3 :}]
Neutron one-particle levels as a function of axially-symmetric 
quadrupole deformation.
Parameters of the Woods-Saxon potential are designed approximately for 
the stable nucleus $^{52}_{24}$Cr$_{28}$.
The diffuseness, the radius and the depth of the Woods-Saxon potential are 
0.67 fm, 4.74 fm, and $-$49.0 MeV, respectively.
See the caption to Fig. 1 and the text for details.
\end{description}

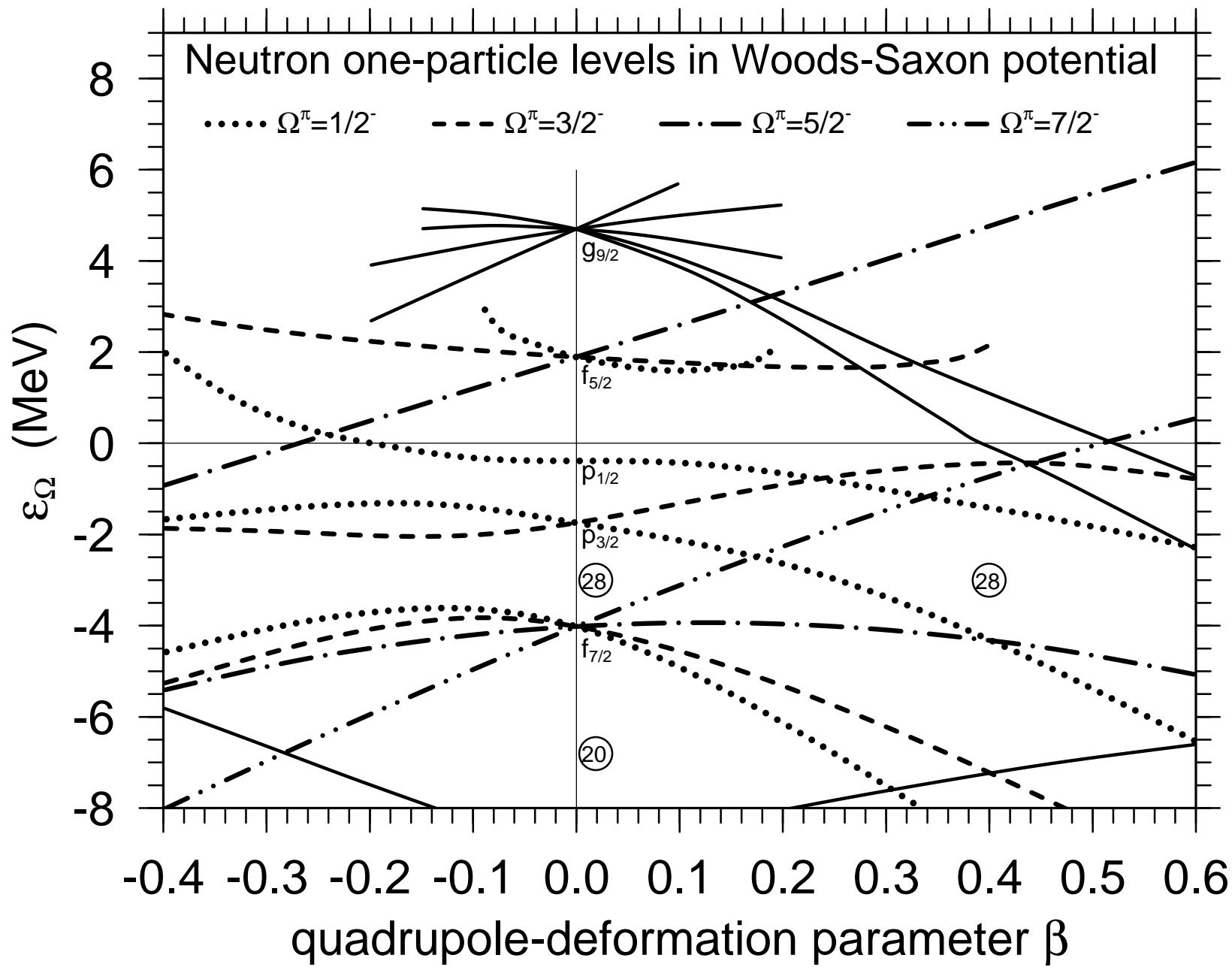
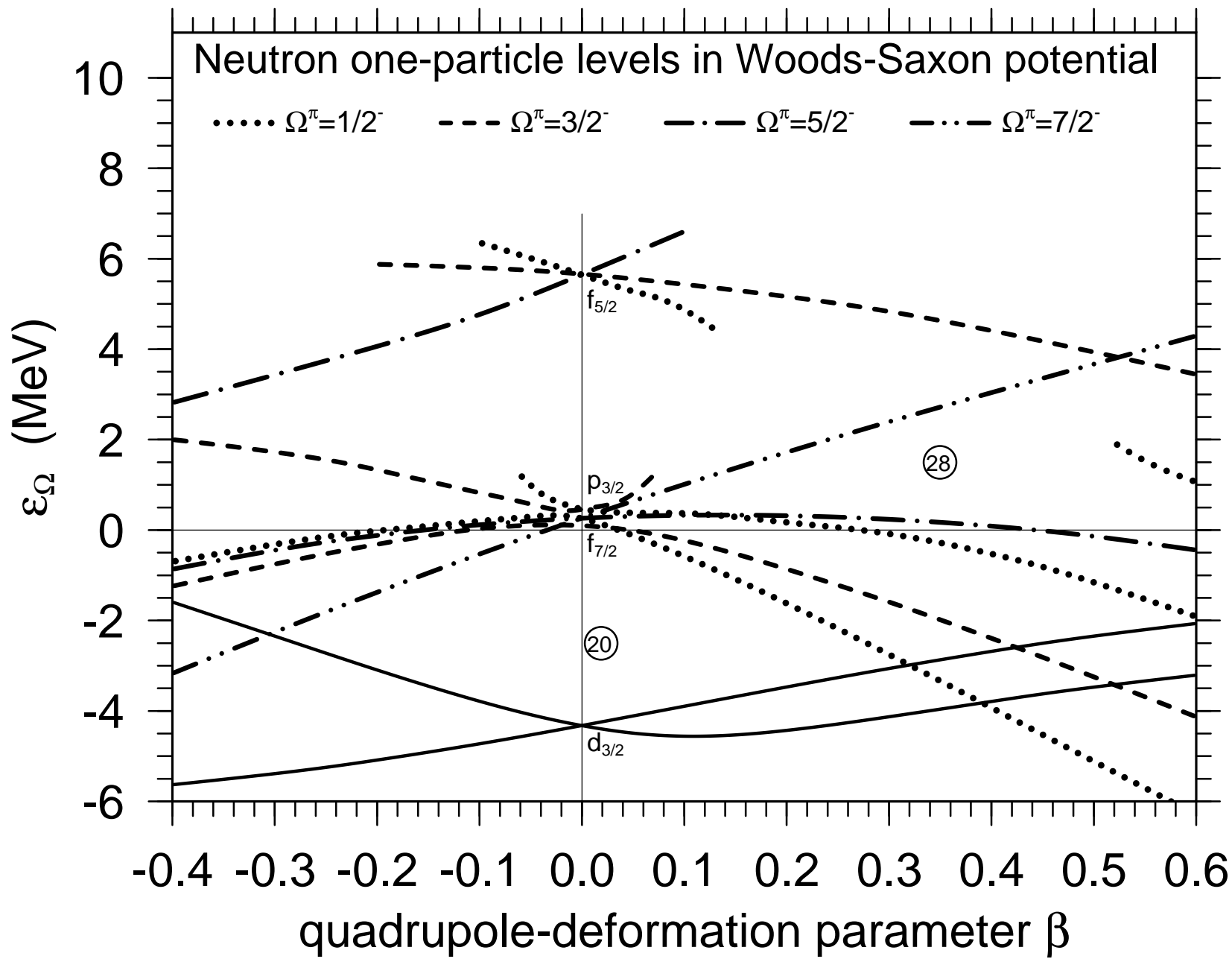
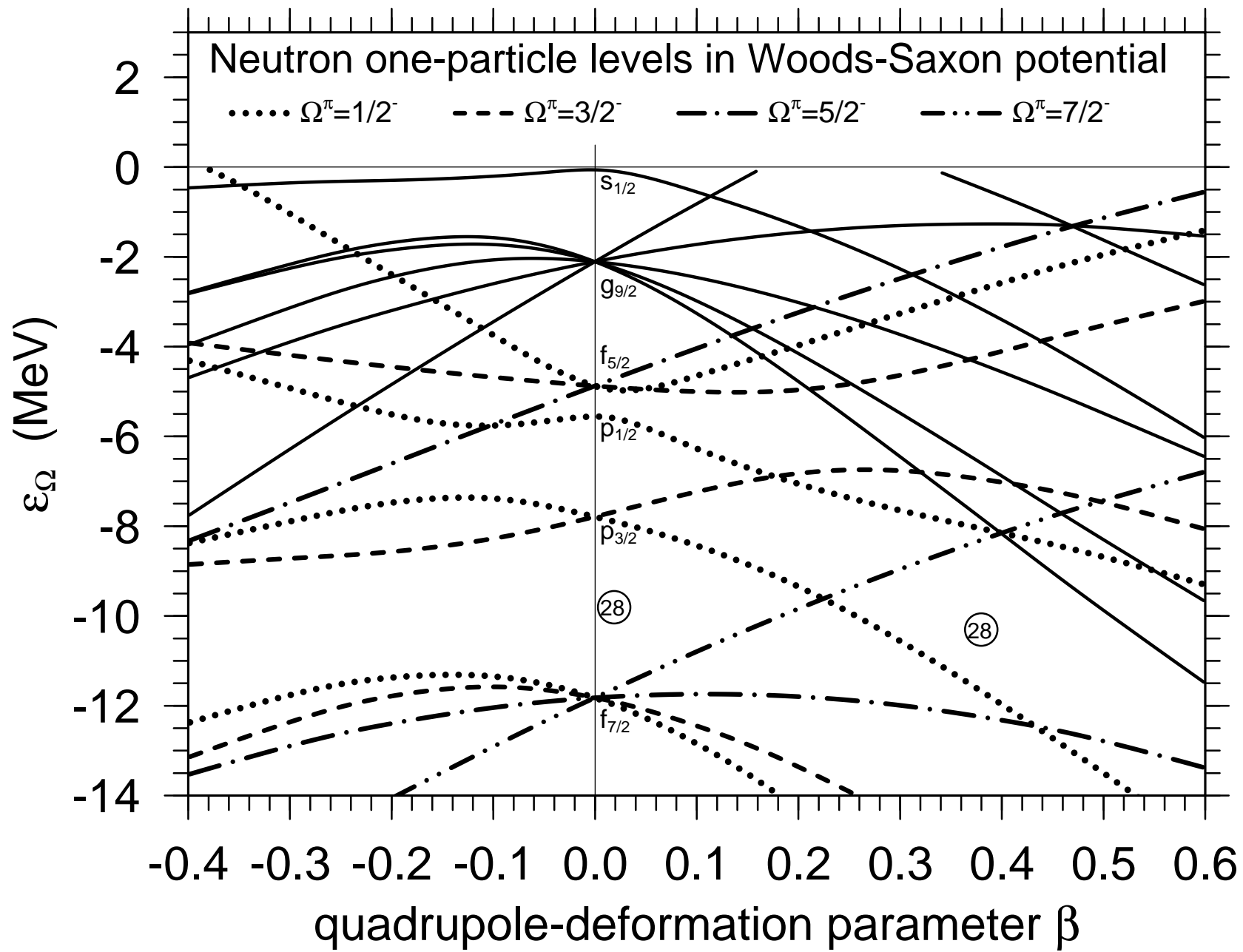
\end{document}